\documentclass[10pt,leqno,titlepage]{amsart}
\usepackage[foot]{amsaddr}
\usepackage{amsmath}
\usepackage{graphicx,psfrag,epsf}
\usepackage{enumerate}
\usepackage{natbib}
\usepackage{url} %

\usepackage{mathtools, empheq}
\usepackage{amssymb, setspace}
\usepackage{bm, bbm}
\usepackage{graphicx, tabularx, array}
\usepackage[svgnames,table]{xcolor}
\usepackage{enumitem}
\usepackage{multirow}
\usepackage{hyperref}
\usepackage{geometry}
\usepackage{makecell}
\usepackage{booktabs}
\usepackage{amsmath}
\usepackage{array}
\usepackage{algorithm}
\usepackage{algorithmic}
\usepackage{tikz}
\usepackage{caption}
\usepackage{float}

\usepackage[normalem]{ulem}

\usepackage{longtable}
\usepackage{threeparttable} %

\usepackage{colortbl} %

\newcolumntype{C}[1]{>{\centering\arraybackslash}m{#1}}

\numberwithin{equation}{section}

\usepackage{amssymb,amsthm,amsmath}

\usepackage{etoolbox}
\AtBeginDocument{%
    \let\orignewpage\newpage 
    \renewcommand\newpage{}
    \patchcmd{\clearpage}{\newpage}{\orignewpage}{}{}}

\begin{document}
\bibliographystyle{plainnat}

\def\spacingset#1{\renewcommand{\baselinestretch}%
{#1}\small\normalsize} \spacingset{1}

\newlist{steps}{enumerate}{1}
\setlist[steps, 1]{label = Step \arabic*:}

\title[ACEs Study Protocol]{PROTOCOL FOR AN OBSERVATIONAL STUDY ON THE EFFECTS OF COMBINATIONS OF ADVERSE CHILDHOOD EXPERIENCES ON ADULT DEPRESSION} 

\author{Ruizhe Zhang$^{1}$}
\author{Jooyoung Kong$^{2}$} 
\author{Dylan S. Small$^{3}$}
\author{William Bekerman$^{3}$}

\thanks{\textbf{Corresponding authors}:\\  William Bekerman, bekerman@wharton.upenn.edu; Ruizhe Zhang, zhangrz22@m.fudan.edu.cn}

\dedicatory{$^{1}$Fudan University, Shanghai, China\\$^{2}$Sandra Rosenbaum School of Social Work, University of Wisconsin-Madison, Madison, WI, USA\\$^{3}$Department of Statistics and Data Science, University of Pennsylvania, Philadelphia, PA, USA}

\begin{abstract}
Adverse childhood experiences (ACEs) have been linked to a wide range of negative health outcomes in adulthood. However, few studies have investigated what specific combinations of ACEs most substantially impact mental health. In this article, we provide the protocol for our observational study of the effects of combinations of ACEs on adult depression. We use data from the 2023 Behavioral Risk Factor Surveillance System (BRFSS) to  assess these effects. We will evaluate the replicability of our findings by splitting the sample into two discrete subpopulations of individuals. We employ data turnover for this analysis, enabling a single team of statisticians and domain experts to collaboratively evaluate the strength of evidence, and also integrating both qualitative and quantitative insights from exploratory data analysis. We outline our analysis plan using this method and conclude with a brief discussion of several specifics for our study.
\end{abstract}

\maketitle

\spacingset{1.5}

\section{Introduction}

Adverse childhood experiences (ACEs) describe specific categories of early life hardship or harm that occur before an individual reaches adulthood. Although these can take various forms, they are typically categorized into ten types involving abuse (physical, emotional, or sexual) and neglect (physical or emotional), as well as household challenges including mental illness, substance abuse, intimate partner violence, parental absence due to divorce or separation, or incarceration. ACEs were initially investigated in a landmark study conducted by the Centers for Disease Control and Prevention (CDC) and Kaiser Permanente, and are highly prevalent and exhibit a strong dose-response relationship with numerous negative health and social outcomes in adulthood (\cite{felitti1998relationship}; \cite{dube2003impact}; \cite{anda2006enduring}).

Exposure to individual ACEs has been repeatedly associated with a range of adverse physical and emotional health consequences later in life. One study has found that childhood emotional abuse increases the risk of lifetime depressive disorders, with adjusted odds ratios (ORs) of 2.7 for women and 2.5 for men (\cite{chapman2004adverse}). Depression also appears to be closely related to having experienced parental alcohol abuse (\cite{anda2002adverse}). Moreover, physical and verbal abuse have been found to be strongly associated with body weight issues and obesity (\cite{williamson2002body}).

Unfortunately, when children are exposed to one type of adversity, there is a high likelihood of exposure to additional ACEs (\cite{dong2004interrelatedness}; \cite{felitti2002relationship}). Numerous large-scale investigations have confirmed that these detrimental experiences can have a cumulative effect: each additional type of adversity elevates the probability of unfavorable outcomes (\cite{felitti1998relationship}; \cite{chapman2004adverse}). Accordingly, previous literature has closely examined the number of categories of ACEs, often referred to as the ACE score. For instance, one study using data from 14 states revealed substantial variation across states in the percentage change in health outcomes for those exposed to at least four ACEs compared to those with fewer than four ACEs (\cite{waehrer2020disease}). Physically, individuals with four or more ACEs face elevated risks of heart disease, liver disease, kidney disease, asthma, Alzheimer’s disease, stroke, metabolic disorders, obesity, diabetes, and certain cancers (\cite{CDC2017}; \cite{CYW2014}; \cite{merrick2019vital}; \cite{hughes2017effect}; \cite{waehrer2020disease}). On the emotional and psychological side, the ACE score strongly correlates with mental health issues such as depression, anxiety, and suicidal ideation (\cite{hillis2004association}; \cite{chapman2004adverse}). Meta-analyses further reinforce that an accumulation of at least 4 ACEs correlates with higher risks of chronic disease, poor mental health, and negative social outcomes, demonstrating the need for strategies to prevent ACEs and lessen their long-term repercussions (\cite{hughes2017effect}).

A growing literature cautions that the additive ACE score is a blunt tool for individual risk, as it treats all events as equal. Then, a child enduring repeated sexual abuse is assumed the same as exposure to parental divorce,  which appears problematic on its face. The ACE score can encourage over-simplified claims about risk or causality, deterministic framing, and stigma (\cite{lacey2020practitioner}). Methodologically, large cohort analyses show that an ACE score cutoff such as $\geq 4$ poorly distinguishes which specific individuals will develop later health problems (\cite{baldwin2021population}; 
\cite{meehan2022poor}). A cohort study of women in correctional custody showed that maltreatment alone is significantly associated with adult mental health while household dysfunction is not. Therefore, the authors challenged the overall ACE score and recommended decomposing the measure back into specific items (\cite{fitzgerald2024challenging}). Recent pediatric guidance finds insufficient evidence to support routine, threshold-based ACE screening in clinical settings and supports trauma-informed, comprehensive assessment instead (\cite{austin2024screening}). Gaps also remain in our understanding of how the accumulation of ACEs influences health. We are therefore interested in which categories of ACEs—and which specific combinations of ACEs (e.g., physical abuse + sexual abuse, or physical abuse + emotional neglect + substance abuse)—are most harmful and warrant particular attention in primary care settings. We focus our analysis on depression, one of the most prevalent and burdensome mental health conditions globally. Consequently, our focus is to evaluate the effect of different combinations of specific ACE items on depression and to identify a higher-risk group of sets of ACEs.

We will examine the impacts of combinations of ACEs on adult depression using data from the Behavioral Risk Factor Surveillance System (BRFSS) 2023, an ongoing health-related telephone survey system administered by the CDC with a random sample of over 400,000 individuals across the United States in 2023. These data include detailed information from ten states that administered the CDC's Adverse Childhood Experiences module in 2023. We observe all ten ACE categories proposed in the CDC-Kaiser study (\cite{felitti1998relationship}), and there are over 50,000 individuals with complete records of all ACE items. Among them, around 17,000 reported zero ACEs while the rest were exposed to at least one ACE. This large sample size will help support powerful data analysis and facilitate the exploration of influencing mechanism of different ACEs combinations.

We employ a novel statistical design called data turnover (Bekerman et al. in preparation, see \cite{bekerman2024protocol} for a description) to conduct this analysis. By integrating both qualitative and quantitative insights from data exploration, data turnover enables a single team of statisticians and domain experts to evaluate evidence from multiple data splits. Exploratory data analysis (EDA) allows us to inspect the data before making potentially questionable assumptions, detect anomalies or outliers that do not represent natural variation in the data, and formulate new hypotheses. It also helps identify variables that may not measure what we initially intended, indicating the need for a better measure or refined, data-informed hypotheses (\cite{bekerman2024planning}). Beyond exploration, replication is critical in observational studies for reinforcing the credibility of findings and sustaining trust in scientific knowledge. Previous work has combined some—but not all—aspects of exploration and replication within the same observational study. For instance, some methods allow for exploration without replication (e.g., \cite{cox1975note}; \cite{heller2009split}), whereas others enable replication without exploration (e.g., \cite{zhao2018cross}; \cite{karmakar2019integrating}). There is also research that supports both exploration and replication but requires two independent investigative teams (\cite{roy2022protocol}). In contrast, data turnover makes it possible to conduct both exploration and replication using a single research group.

In observational studies like ours, where treatment assignment is not under our direct control, obtaining consistent findings across multiple groups that may have different assignment mechanisms reinforces the evidence that the treatment itself is truly responsible for the observed effects (\cite{rosenbaum2001replicating}; \cite{rosenbaum2015see}). By analyzing data from two distinct subpopulations of BRFSS 2023 participants—residing in states that voted for the Democratic candidate in the 2024 US presidential election ("blue states") and those residing in states that voted for the Republican candidate ("red states") —we can assess the replicability of our findings. We split our data into blue states and red states, recognizing that these political orientations often shape distinct social, economic, and health care policies. A study focusing on the population in East and West Germany has shown that socio-political context is a determinant of childhood maltreatment (\cite{ulke2021socio}). In our context, for example, blue states may allocate more resources toward public health programs (e.g., adopt Medicaid expansion) or have broader social safety nets, whereas red states may emphasize different strategies for healthcare funding and child protective services. Such policy and cultural variations can influence both the prevalence of ACEs and the mechanisms by which individuals are exposed to them, potentially resulting in different patterns of ACEs across these two groups. Demonstrating consistent results in both blue and red states thus enhances confidence that our findings are robust across varied socio-political contexts. By employing data turnover, we can predetermine how one subgroup will guide analysis of the other, then apply exploratory data analysis to the second subgroup to inform how we draw inferences about the first. Any outcome that shows a significant effect in both groups is defined as a “replicable” finding (\cite{bogomolov2023replicability}). In Section \ref{analysis}, we detail how we apply this non-random cross-screening method to detect replicable outcomes and test the global null hypothesis, i.e., identifying outcomes that are affected in at least one of the two subpopulations.

The remainder of the document is organized as follows. We introduce the BRFSS dataset and describe our data processing in Section \ref{brfss}. In Section \ref{iss}, we outline our methodology. Finally, we conclude with a description of our data analysis procedure and an explanation of several specifics of implementation in Section \ref{analysis}.

\section{The Behavioral Risk Factor Surveillance System (BRFSS)} \label{brfss}

This study uses data from BRFSS, an ongoing health-related telephone survey system administered by the CDC. BRFSS collects comprehensive data on a wide range of health-related behaviors, chronic health conditions, healthcare access, and preventive service usage from the non-institutionalized adult population (18 years and older). The survey employs a dual-frame sampling methodology, using both landline and cellular telephone surveys to ensure representativeness of the data. The dataset collected in 2023 is the latest data available, has 433,323 observations from 48 states, the District of Columbia, Guam, Puerto Rico, and the US Virgin Islands, and includes topics such as health status, chronic conditions, health behaviors, healthcare access, and preventive services, as well as optional modules including social determinants of health, cancer screenings, and cognitive decline.

For the purpose of our study, we focus on the 10-state\footnote{Delaware, Florida, Georgia, Missouri, Nevada, New Jersey, Oregon, Rhode Island, Tennessee, Virginia.} sample that administered the CDC's ACEs module in 2023.  Following previous studies utilizing the BRFSS data (\cite{bhan2014childhood}; \cite{gilbert2015childhood}; \cite{waehrer2020disease}), we collapse different questions into binary indicators of each ACE. More details are provided in Table \ref{tab:data_processing}. In total, we create 10 binary ACE items from the BRFSS data (see Table \ref{tab:ace_items_brfss}), which is consistent with the categorization in the CDC-Kaiser study (\cite{felitti1998relationship}). Depression is coded $Y{=}1$ if the answer to \emph{'(Ever told) (you had) a depressive disorder (including depression, major depression, dysthymia, or minor depression)?'} is \emph{'Yes'} and $0$ otherwise.

\begin{table}[h]
\centering
\begin{tabular}{|>{\centering\arraybackslash}m{4cm}|>{\centering\arraybackslash}m{6cm}|>{\centering\arraybackslash}m{4cm}|}
\hline
\textbf{ACEs Item} & \textbf{Questions Collapsed} & \textbf{Encoding Method} \\ \hline
Any Sexual Abuse & Number of times forced to touch, be touched, or have sex with anyone at least five years older or an adult & Constructed by summing responses across multiple questions: all “never” encoded as 0, otherwise encoded as 1  \\ \hline
Household Substance Use & Alcohol and illegal drug use in the household & Constructed by summing responses across multiple questions: all “never” encoded as 0, otherwise encoded as 1 \\ \hline
Physical Abuse & Frequency of physical abuse & ‘Once’ or ‘More than once' encoded as 1, ‘None’ encoded as 0 \\ \hline
Verbal Abuse & Frequency of verbal abuse & ‘Once’ or ‘More than once' encoded as 1, ‘None’ encoded as 0 \\ \hline
Parental Divorce & Response to parental divorce question & 'Yes' encoded as 1, 'No' encoded as 0 \\ \hline
Living with a Depressed or Mentally Ill Person & Response to question about living with a mentally ill person & 'Yes' encoded as 1, 'No' encoded as 0 \\ \hline
Living with an Incarcerated Person & Response to question about living with an incarcerated person & 'Yes' encoded as 1, 'No' encoded as 0 \\ \hline
Physical Violence Between Parents & Frequency of physical violence between parents & ‘Once’ or ‘More than once' encoded as 1, ‘None’ encoded as 0 \\ \hline
Emotional Neglect & For how much of your childhood was there an adult in your household who made you feel safe and protected & ‘Never’ and ‘A little of the time’ encoded as 1, ‘Some of the time’, ‘Most of the time’, ‘All of the time’ encoded as 0 \\ \hline
Physical Neglect & For how much of your childhood was there an adult in your household who tried hard to make sure your basic needs were met & ‘Never’ and ‘A little of the time’ encoded as 0, ‘Some of the time’, ‘Most of the time’, ‘All of the time’ encoded as 1 \\ \hline
\end{tabular}
\caption{Data Processing Steps and Collapsed Indicators}
\label{tab:data_processing}
\end{table}

\begin{table}[h]
\centering
\begin{tabular}{ll}
  \hline
  \textbf{ACE Item} & \textbf{Variable} \\
  \hline
  Live With Anyone Depressed, Mentally Ill, Or Suicidal & ACEDEPRS \\
  Substance Abuse in the Home & ACESUB \\
  Live With Anyone Who Served Time in Prison or Jail & ACEPRIS \\
  Were Your Parents Divorced/Separated & ACEDIVRC \\
  How Often Did Your Parents Beat Each Other Up & ACEPUNCH \\
  How Often Did A Parent Physically Hurt You In Any Way & ACEHURT1 \\
  How Often Did A Parent Swear At You & ACESWEAR \\
  Sexual Abuse & ACESEX \\
  Did An Adult Make You Feel Safe And Protected & ACEADSAF \\
  Did An Adult Make Sure Basic Needs Were Met & ACEADNED \\
  \hline
\end{tabular}
\caption{ACE Items in BRFSS 2023}
\label{tab:ace_items_brfss}
\end{table}

After processing the data, we restrict our sample to those who do not have any missing values in any of the ten ACE variables. We also omit those individuals who do not have the appropriate measurement for depression. In total, there are 49,547 remaining individuals, among whom 23,390 individuals are from blue states\footnote{Delaware, New Jersey, Oregon, Rhode Island, Virginia} (as established in the 2024 United States presidential election) and 26,157 individuals from red states\footnote{Florida, Georgia, Missouri, Nevada, Tennessee}. In the part of data with individuals from blue states, 7,363 individuals do not have any ACEs while 16,027 were exposed to at least one ACE. In the part of data with individuals from red states, 8,479 individuals do not have any ACEs while 17,678 were exposed to at least one ACE.

\section{Isotonic Subgroup Selection}
\label{iss}

We plan to use isotonic subgroup selection (ISS) to identify the high-risk group in our study (\cite{muller2025isotonic}). Let \((X_i, Y_i)\), for \(i = 1, \dots, n\), be independent observations, where \(X_i \in \mathbb{R}^d\) denotes covariates and \(Y_i\) is the outcome of interest. In our setting, \(Y_i\) is a binary indicator for depression, and \(X_i\) consists of different combinations of ten ACE variables, each coded as binary. We assume a regression function $\eta(x) = \mathbb{E}(Y|X = x)$ and impose the coordinate-wise monotonicity assumption: $x \preceq x' \implies \eta(x) \le \eta(x')$, where $x \preceq x'$ means $x_j \le x'_j$ for all coordinates $j = 1, \dots, d$. In practice, this monotonicity captures the expectation that exposure to additional ACEs should not reduce the probability of depression. For example, an individual reporting both physical abuse and parental substance use is assumed to have risk at least as high as someone reporting only parental substance use.

Fix a threshold $\tau$. Our target is the superlevel region $X_\tau(\eta) = \{\,x : \eta(x) \ge \tau\,\}$, which we interpret as the subgroup of covariate profiles whose expected probability of depression exceeds $\tau$. The goal of Isotonic Subgroup Selection (ISS) is to estimate a subset $\hat A$ that approximates $X_\tau(\eta)$ while controlling error rates in a principled way.

The ISS procedure consists of three main steps: 
\begin{enumerate}[topsep=2pt,itemsep=2pt,parsep=0pt,partopsep=0pt]
  \item \textbf{Hypothesis construction.} 
  
  For each $X_i$, we test $H_{0,i}: \eta(X_i) < \tau \quad\text{versus}\quad H_{1,i}: \eta(X_i) \ge \tau.$
  Because $Y$ is binary, classification-based $p$-values are constructed using anytime-valid martingale methods for bounded outcomes. Intuitively, these $p$-values are based on the cumulative evidence from neighboring profiles $X_j \preceq X_i$, comparing the observed number of depressed cases to what would be expected if $\eta(X_i) < \tau$. This ensures that even with dependent tests across profiles, each $p_i$ is valid.

  Formally, let $I(x) = \{\,j : X_j \preceq x\,\}$, $n(x) = |I(x)|$, and denote the concomitant responses by $Y_{(1)}(x), \dots, Y_{(n(x))}(x)$. Define $S_k(x) = \sum_{j=1}^k Y_{(j)}(x)$. Then an anytime-valid $p$-value for testing $\eta(x) < \tau$ is
  $$
    p_\tau(x) = \min_{1 \le k \le n(x)} 
      \frac{\tau^{\,S_k(x)} (1-\tau)^{\,k - S_k(x) + 1}}
           {B(1-\tau;\, k - S_k(x) + 1,\, S_k(x) + 1)},
  $$
  where $B(z;a,b)$ is the incomplete Beta function. This construction ensures that $\mathbb{P}(p_\tau(x) \le \alpha \mid X) \le \alpha$ whenever $\eta(x) < \tau$ (see \cite{muller2025isotonic}, Lemma 22).

  \item \textbf{Logical structure and multiple testing.} 
  
  The set of hypotheses $\{H_{0,i}\}$ is not independent. By monotonicity, if $X_i \preceq X_j$, then $H_{0,j} \subseteq H_{0,i}$. To encode these logical relations, we construct a directed acyclic graph (DAG): $G = (I,E)$, where $I=\{1,\dots,m\}$ indexes the hypotheses. We include a directed edge $i \to j$ whenever $X_j \preceq X_i$, so that edges always point from a stronger profile to a weaker one. From $G$, we derive a polyforest $F$ by retaining at most one parent for each node. This choice is not unique and can be made, for example, at random, by selecting the closest parent in $\ell_\infty$ norm, or alternatively using other distance measures such as the $\ell_2$ norm. This sparsification reduces complexity while preserving logical consistency.

  Let $R^\text{ISS}_\alpha(G,p)$ denote the rejection set from the DAG testing algorithm. The procedure begins by distributing the global $\alpha$-budget across the root nodes of $F$, with each root $i$ receiving share
  $$
    \alpha_i = \frac{\#\{\text{leaf descendants of } i \text{ in } F\}}{\#\{\text{all leaf nodes in } F\}} \cdot \alpha.
  $$
  A root node $i$ is rejected if $p_{\tau,i} \le \alpha_i$, in which case all its ancestors in $G$ are also rejected. It's worth noting that here all its ancestors are defined with respect to the full DAG $G$, not only within the same tree of the polyforest. Any node that is logically stronger (i.e., dominates $i$ under the partial order) is simultaneously rejected. After each iteration, the rejected nodes are removed from $F$, and the unused $\alpha$ is reallocated to the remaining roots. The process continues until no further rejections are possible. (see formal pseudocode in Algorithm 1, \cite{muller2025isotonic})

  Intuitively, the algorithm passes significance level downward through the hierarchy, where roots receive budget first, and rejected nodes release unused $\alpha$ to their descendants. Ancestors are rejected whenever a descendant is, ensuring monotonicity. The final rejection set $R^\text{ISS}_\alpha(G,p)$ therefore respects both the partial order and the global FWER.

  \item \textbf{Upward closure.} 
  
   Denote the set of rejected nodes from the polyforest testing procedure as $R_\alpha^\text{ISS}(G,p)$. The estimated subgroup is then defined as $\hat A = \{\,x : x \succeq X_i \text{ for some } i \in R_\alpha^\text{ISS}(G,p)\,\}.$
  By construction, $\hat A$ is the upward closure of $R_\alpha^\text{ISS}(G,p)$ with respect to the partial order $\preceq$. This closure step ensures that the selected subgroup respects the monotonicity structure: once a particular exposure profile $X_i$ is identified as high-risk, all more severe profiles $x \succeq X_i$ must also be included. 
\end{enumerate}
Under this construction, ISS guarantees that with probability at least $1-\alpha$, no profile with $\eta(x)<\tau$ is included in $\hat A$, controlling the family-wise error rate (FWER). In applied terms, ISS provides a conservative but rigorous way to delineate ACE exposure patterns linked to elevated depression risk. Concretely, this means that the subgroups we identify consist only of ACE combinations whose estimated probability of depression exceeds the chosen threshold, while still capturing as many truly high-risk profiles as possible. In this way, ISS leverages both valid statistical guarantees and the monotonicity structure to yield interpretable subgroups of adverse childhood experiences most strongly associated with depression.

\section{Design of Data Analysis} \label{analysis}
We aim to explore the effect of different combinations of ACEs on depression and also identify those with higher risk. We will test all of the possible combinations of ACEs, therefore leading to 1024 ($2^{10}$) different hypotheses. We will deem those combinations of ACEs for which the effect more than doubles the odds of depression in both subgroups - those residing in blue states and red states - as replicable findings. Also, we are interested in those combinations more than doubling the odds ratio of depression in only one subgroup, that is testing the global null hypothesis.

\subsection{Data Turnover Framework} 

We employ data turnover (\cite{bekerman2024protocol}) to analyze the data, ensuring control of the family-wise error rate (FWER; the chance of falsely rejecting any true null hypotheses at level $\alpha=0.05$) and facilitating flexible inference compared to automated cross-screening (\cite{zhao2018cross}), by integrating both qualitative and quantitative insights from data exploration and formulating data-informed new hypotheses (e.g., \cite{bekerman2024planning}).

A brief description is as follows. We split our data into two subgroups: those individuals in our study residing in blue states and those residing in red states. The first subgroup, without data exploration, suggests a limited number of combinations of ACEs to be validated in the second subgroup based on some pre-determined selection procedure. Then, we do \textit{ad hoc} data exploration on the second subgroup to choose how to conduct one-sided hypothesis tests on the first subgroup, possibly allowing for novel hypotheses to be identified and tested. In other words, both subgroups are used to screen and both subgroups are used to validate, but only one subgroup can generate novel hypotheses to be validated in the other subgroup while still maintaining the validity of our entire procedure. Because a limited number of combinations seek validation on the second subgroup, compared with directly controlling the FWER of 1024 hypotheses on the whole data (eg. by using Bonferroni correction, Holm procedure or directly applying ISS on the whole sample), only a small fraction of the 1024 potential hypotheses are considered in the validation sample. 

Moreover, we need to determine whether it is more beneficial to conduct EDA on the smaller or larger sample. Simulation results in Bekerman’s paper (\cite{bekerman2024protocol}) suggest that performing EDA on the smaller subgroup—in our case, participants living in blue states—is advantageous. We anticipate that incorporating data exploration within this subgroup will strengthen our study’s overall design by enabling data-informed decision-making and facilitating the emergence of new hypotheses.

\subsection{Analysis Plan}

As we note before, for each $X_i$, we test $H_{0,i}: \eta(X_i) < \tau \quad\text{versus}\quad H_{1,i}: \eta(X_i) \ge \tau. $ Therefore, we need to
prespecify $\tau$. Intuitively, if $\tau$ is too small, the true superlevel set is very large and its corners tend to lie deep. If $\tau$ is too large, no method has power and all outputs are empty. So we are interested in testing whether a combination of ACEs would more than double the odds of depression compared with the reference population. Let $p_0$ be a baseline probability - proportion of people exposed to no ACEs. In previous BRFSS datasets which are not included in this study, $p_0$ is around $9.4\%$. To test that the odds at $x$ are at least $c$ times the baseline odds, \[\frac{\tau}{1-\tau}=c \frac{p_0}{1-p_0}\]
\[\tau\ = \frac{cp_0}{1+(c-1)p_0}\]
Use $c=2, p_0=9.4\%$, $\tau=0.172$.

We plan to integrate isotonic subgroup selection (ISS) with the data turnover method. Recall that in data turnover framework, we split the data into two subgroups $\mathcal{I}_R$ (Red Subgroup) and $\mathcal{I}_B$ (Blue Subgroup). 
For every combination $X_i$ that we are interested, define the anytime-valid classification $p$-value $p_\tau(x)$ as in the binary setting (see Section~\ref{iss}). 
Below is the pre-specified analysis plan for the red subgroup. Compute red-side $p$-values $p^{R}_{\tau,i}:= p_\tau(X_i;\mathcal{I}_R)$ for all $i$, and form the screened index set $\mathcal{S}_\kappa = \bigl\{\, i : p^{R}_{\tau,i} \le \kappa \,\bigr\}$ to be tested in $\mathcal{I}_B$, for a pre-specified screening cutoff $\kappa \in (0,1)$. From the full DAG $G$, we derive a polyforest $F$ by retaining at most one parent for each node, which can be made at random or based on a distance criterion. We could organize the polyforest $F_\kappa^B$ in $\mathcal{I}_B$ from the information ($p^{R}_{\tau,i}$, $\mathcal{S}_\kappa$) we obtain from the screening stage in $\mathcal{I}_R$. All nodes of the polyforest $F_\kappa^B$ are drawn from \(\mathcal{S}_\kappa\).
For each $i\in\mathcal{S}_\kappa$, define its set of \textit{covers}
$$
\mathrm{Cover}(i) \;=\; \bigl\{\, j\in \mathcal{S}_\kappa \,:\, X_j \succ X_i \ \text{and there is no } k\in \mathcal{S}_\kappa \text{ with } X_j \succ X_k \succ X_i \,\bigr\}.
$$
Select at most one parent for $i$ from $\mathrm{Cover}(i)$ guided by $p$-value:
$\mathrm{par}(i) \in 
\arg\min_{\,j \in \mathrm{Cover}(i)} \ p^{R}_{\tau,j}$, choosing among candidates with tied minimal p-value uniformly at random. If $\mathrm{Cover}(i)=\emptyset$, declare $i$ a root. 
This yields the polyforest $F_\kappa^B$ (a DAG with in-degree $\leq 1$). 
On $F_\kappa^B$, compute blue-side $p$-values only for screened nodes, $ p^{B}_{\tau,i}:= p_\tau(X_i;\mathcal{I}_B)$ for $i\in\mathcal{S}_\kappa$, and run the ISS DAG-testing algorithm with polyforest $F_\kappa^B$ controlling the FWER for the blue part at level $\alpha/2$. 

Using $p_{\tau,i}^{R}$ to choose parents steers each node toward a path whose upstream evidence is strongest, so $\alpha$ is concentrated along promising chains. 
This mitigates cases where a large-$p$ maximal node would otherwise block the flow of $\alpha$ and prevent strong descendants from ever receiving budget, increasing the power. 

\begin{figure}[H]
    \centering
    \includegraphics[width=1\linewidth]{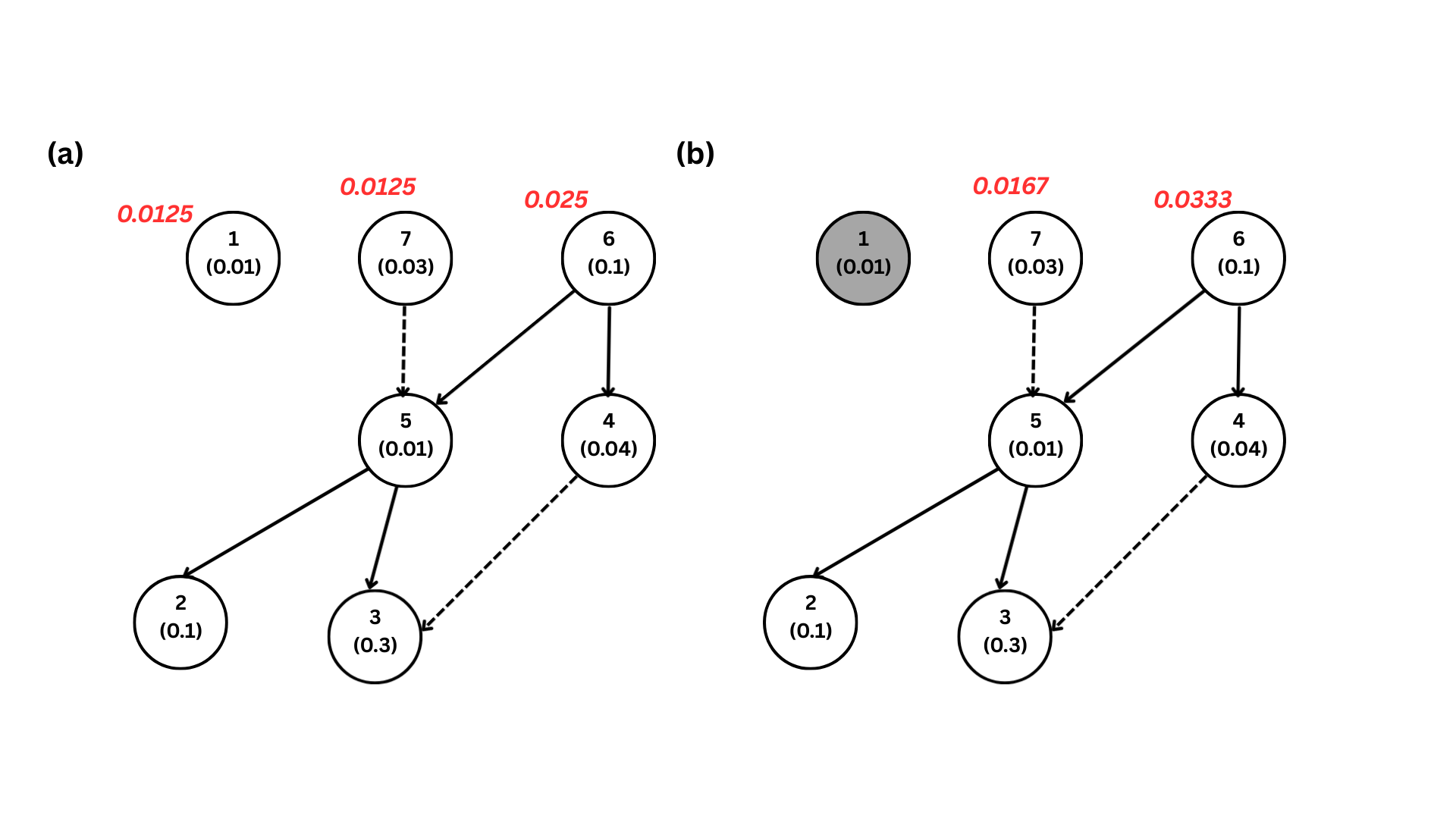}
    \caption{DAG Testing: Example 1. Each node is labeled with its $p$-value (round brackets); in the induced polyforest-weighted DAG, solid arrows indicate edges that are retained in the polyforest (weight$=$1), while dashed arrows indicate edges that exist in the full DAG but are not retained in the polyforest (weight$=$0, since each node can have at most one parent); filled circles indicate hypotheses previously rejected. Panels (a)–(b) correspond to successive iterations of the algorithm, showing how rejections propagate through the structure.}
    \label{fig:dag eg1}
\end{figure}

Figure \ref{fig:dag eg1} illustrates a case where the parent selection creates a bottleneck. 
Here, nodes $1$, $6$, and $7$ are roots, each receiving part of the global $\alpha$-budget (shown in red). 
Although node $5$ has a small $p$-value ($0.01$), its designated parent is node $6$, whose $p$-value is relatively large ($0.1$). 
Since node $6$ is unlikely to be rejected with its assigned budget ($0.025$), the rejection cannot propagate to node $5$, and the signal at node $5$ is blocked. 
This demonstrates how attaching a promising child to an unpromising parent can prevent significant nodes from being discovered.

\begin{figure}[H]
    \centering
    \includegraphics[width=1\linewidth]{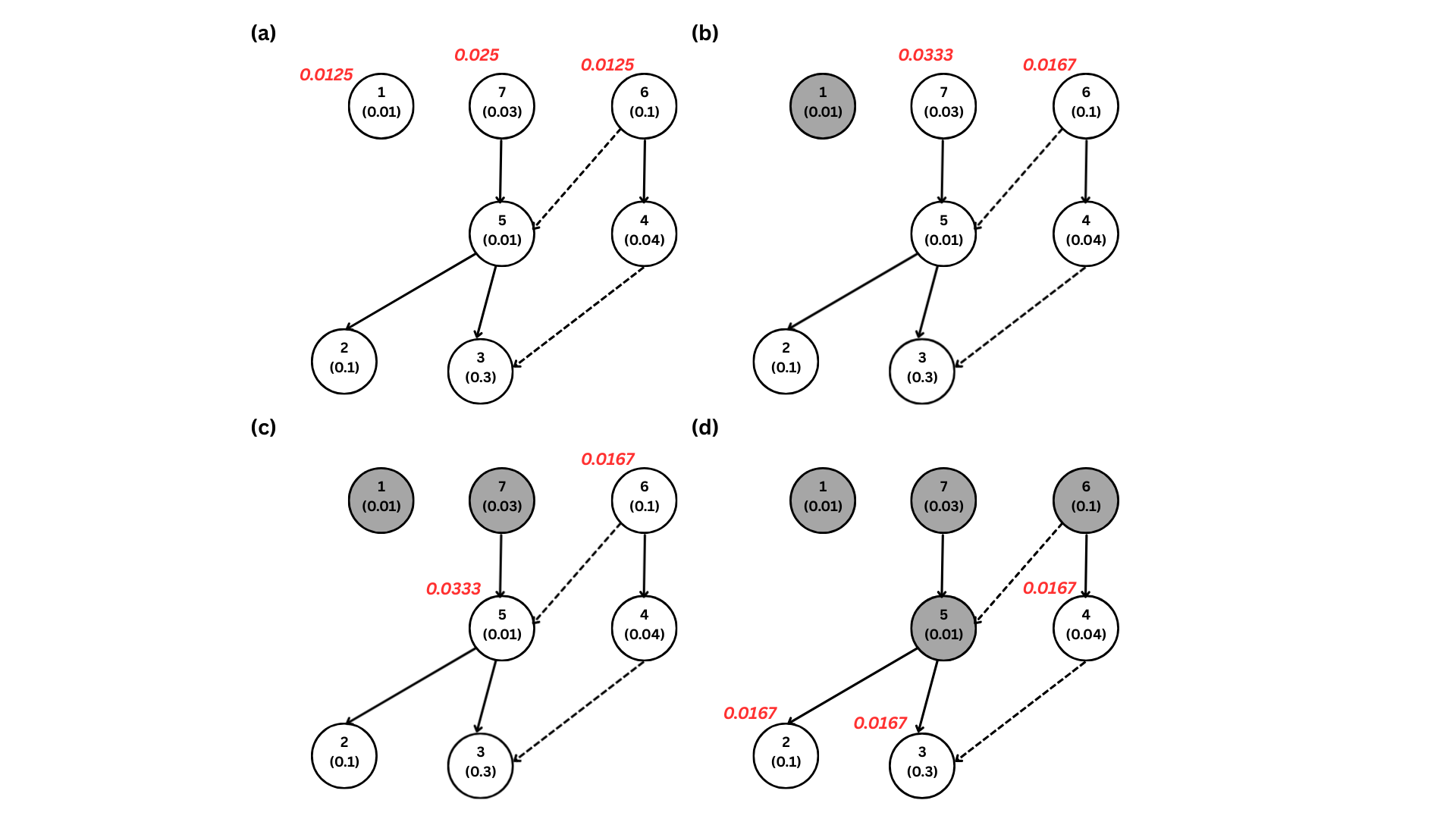}
    \caption{DAG Testing: Example 2. The setting is the same as in Figure \ref{fig:dag eg1}, except that node $5$ is attached through node $7$ rather than node $6$. Panels (a)–(d) correspond to successive iterations of the algorithm, showing how rejections propagate through the structure.}
    \label{fig:dag eg2}
\end{figure}

In contrast, Figure \ref{fig:dag eg2} shows the situation when node $5$ is attached through node $7$, whose $p$-value is smaller ($0.03$). 
After node $1$ is rejected in the first iteration ($p_1=0.01 \leq 0.0125$), the unused budget is reallocated, giving node $7$ a larger share ($0.0333$). 
With $p_7=0.03$, node $7$ can then be rejected, releasing further budget to its child node $5$. 
Since node $5$ has $p_5=0.01$, it is subsequently rejected, and the rejection can continue to its descendants. Simultaneously, since node $6$ is the parent of node $5$ in the full DAG, even if its $p$-value is large, it can also be rejected in line with logical partial order.
Compared to Figure \ref{fig:dag eg1}, the $p$-guided parent selection in Figure \ref{fig:dag eg2} avoids the bottleneck at node $6$ and allows the procedure to detect the significant signal at node $5$.

In the screening stage, we also need to define the cutoff $\kappa$ of p-value (eg. 0.05). In a paper defining multimorbidity (\cite{silber2018defining}), they chose to form a 99.9\% confidence interval in checking if the interval for the odds ratio was completely above 2. This is a stringent threshold to control for the number of hypotheses falsely passing the screening step. Yet, choosing this cutoff is rather heuristic given the sample sizes and number of hypotheses. A basic intuition would be: if the cutoff is too large, then almost all of the hypotheses would pass the screening stage, which would make this step meaningless; if the cutoff is too small, though only a small number of true nulls would falsely pass the screening stage, some true non-nulls would be screened out if the signals aren’t that strong. In our testing procedure, when a root node is rejected, all its ancestors defined with respect to the full DAG are rejected. That is, even if the combination of exposures is logically stronger but yields a comparatively large $p$-value, while a weaker combination produces a small $p$-value and is rejected, the stronger combination will also be rejected. For example, suppose the subgroup defined by “parental substance use only” has $p=0.0001$ and is rejected at its $\alpha$ budget level, whereas the larger subgroup “parental substance use and physical abuse” has $p=0.40$. Because the second subgroup dominates the first in the partial order, it is rejected automatically once the first is, regardless of its own $p$-value. Thus, in our context, the cutoff $\kappa$ could be relatively more lenient. We intend to control the FWER at level $\alpha=0.05$. Therefore, we will apply the DAG-testing procedure on each list of p-values of selected hypotheses at level $\alpha/2=0.025$. Accordingly, we choose 0.025 as our cutoff in the screening stage. In other words, if the anytime-valid p-value of one hypothesis is smaller than 0.025, then this hypothesis will pass the screening step, otherwise, this will be screened out. 

For the blue subgroup - the one with smaller sample size, we will do exploratory data analysis (EDA). Of course, we could first do the pre-specified thing we plan on the red subgroup, but we have more flexibility on this part. This flexibility may allow us to identify more complex hypotheses or yield unexpected benefits. For example, we may explore alternative ways of grouping ACE categories and refine our criteria for encoding ACE items.

\newpage

\bibliography{references.bib}

\orignewpage

\orignewpage

\end{document}